\documentclass[runningheads,a4paper]{llncs}
\usepackage{makeidx}  

\usepackage{upgreek} 
\usepackage{siunitx}
\usepackage{hyperref}
\usepackage{amssymb}
\usepackage{braket}
\usepackage{colonequals}
\usepackage{amsmath}
\usepackage{graphicx}
\usepackage{array}
\newcommand\dif{\mathop{\rm d\!}}
\newcommand\hbu{\hat{\boldsymbol{u}}}

\usepackage{color} 
\definecolor{blue}{rgb}{0,0,0.8}

\definecolor{red}{rgb}{0.8,0,0}

\definecolor{green}{rgb}{0,0.4,0}

\DeclareMathOperator{\Tr}{Tr}

\begin{document}
\title{Optimal experimental design for biophysical modelling in multidimensional diffusion MRI}
\author{Santiago Coelho\inst{1} \and Jose M. Pozo\inst{1}\and Sune N. Jespersen\inst{2,3} \and Alejandro F. Frangi\inst{1} }

\institute{Centre for Computational Imaging \& Simulation Technologies in Biomedicine (CISTIB), School of Computing \& School of Medicine, University of Leeds, Leeds,  UK \email{\{s.coelho,j.m.pozo,a.frangi\}@leeds.ac.uk}
\and Center of Functionally Integrative Neuroscience (CFIN) and MINDLab, Department of Clinical Medicine, Aarhus University, Aarhus, Denmark \email{sune@cfin.au.dk}
\and Department of Physics and Astronomy, Aarhus University, Aarhus, Denmark}
\maketitle
\begin{abstract}


Computational models of biophysical tissue properties have been widely used in diffusion MRI (dMRI) research to elucidate the link between microstructural properties and MR signal formation. For brain tissue, the research community has developed the so-called \textit{Standard Model} (SM) that has been widely used. However, in clinically applicable acquisition protocols, the inverse problem that recovers the SM parameters from a set of MR diffusion measurements using pairs of short pulsed field gradients was shown to be ill-posed. 
Multidimensional dMRI was shown to solve this problem by combining linear and planar tensor encoding data. Given sufficient measurements, multiple choices of b-tensor sets provide enough information to estimate all SM parameters. However, in the presence of noise, some sets will provide better results. In this work, we develop a framework for optimal experimental design of multidimensional dMRI sequences applicable to the SM. This framework is based on maximising the determinant of the Fisher information matrix, which is averaged over the full SM parameter space. This averaging provides a fairly objective information metric tailored for the expected signal but that only depends on the acquisition configuration. The optimisation of this metric can be further restricted to any subclass of desirable design constraints like, for instance, 
hardware-specific constraints. In this work, we compute the optimal acquisitions over the set of all b-tensors with fixed eigenvectors.

\end{abstract}


\section{Introduction}

Diffusion MRI (dMRI) is sensitive to the random displacement of water molecules within a voxel. This allows probing tissue at scales considerably smaller than the image resolution \cite{KISELEV2016}, 
enabling voxel averaged microstructural changes to be monitored with dMRI. 
The ability to detect small alterations in brain tissue is a key factor when developing biomarkers for early stages in neurodegeneration \cite{ASSAF2008b}. The desire to obtain not only sensitive but also specific characterisation of microstructure motivated the development of biophysical tissue models, which can capture subtle changes in tissue microstructure \cite{ASSAF2004a}. The \textit{Standard Model} (SM) has emerged in brain tissue \cite{NOVIKOV2018} as 
an overarching term for a class of previously used similar models. 
However, it has been demonstrated that with conventional multi-shell dMRI acquisitions (linear tensor encoding, LTE) available 
in clinical settings, the estimation of SM parameters is ill-posed \cite{JELESCU2015b}. 
Recent work has shown it is possible to make this estimation problem well-posed by adding functionally independent dMRI measurements such as planar tensor encoding (PTE) data \cite{COELHO2019a,REISERT2019}. However, LTE and PTE are particular cases in the b-tensor space accessible with multidimensional dMRI, and SM parameters can also be unambiguously estimated by other combinations. This begs the question: How should we sample the b-tensor space to minimise the error in the estimated SM parameters?


This work proposes a framework for optimal design of multidimensional dMRI acquisition protocols \cite{WESTIN2016} to estimate biophysical models. We first define a metric based on the \textit{Fisher Information Matrix} (FIM), which quantifies the expected amount of information on a certain b-tensor set (B-set, \textit{i.e.} a set of measurements). We apply such framework to minimise the error in the estimation of the tensors in the fourth order cumulant expansion. This works as a surrogate of the fully unconstrained SM parameters, but has the advantage of providing a general result, potentially applicable to other models containing fewer assumptions. 
The capabilities and feasibility of the proposed framework is tested by computing the optimal B-set for two predefined constraints 
with different degrees of freedom in the optimisation. Numerical experiments show the optimal B-sets have interesting non-trivial distributions and lead to reduced estimation errors in both the cumulant expansion and the SM.


\section{Theory}

\begin{figure}[htbp]
\centerline{\includegraphics[scale=.35]{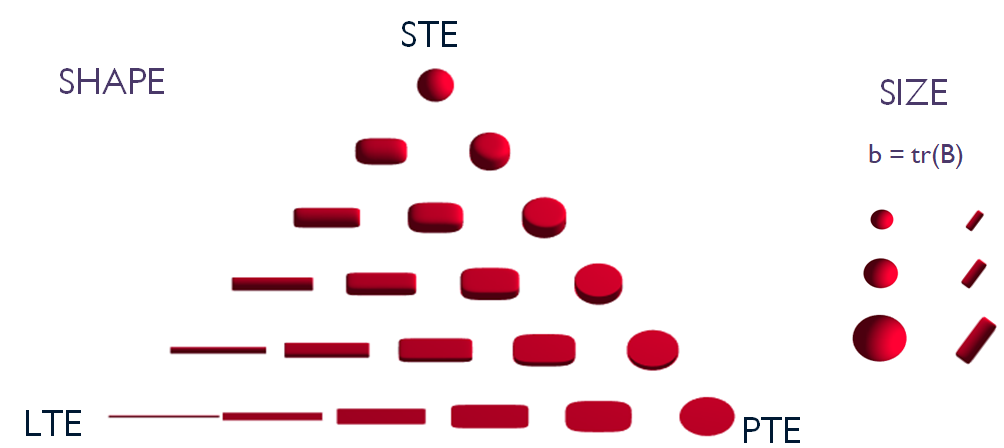}}
\caption{Superquadric tensor glyphs arranged in a barycentric ternary diagram \cite{TOPGAARD2017} according to their linear, planar, and spherical components (LTE, PTE, and STE). Two degrees of freedom define the tensor shape, and an extra one is needed for its size.}
\label{fig:Tensor_Shapes}
\end{figure}

Unlike conventional dMRI acquisitions performed in an LTE framework, a single multidimensional dMRI measurement does not probe a point but a trajectory in q-space. This generalises the concept of diffusion weighting along a direction (b-vector), to more complex scenarios, viz. simultaneously sensitising the MR signal to diffusion along multiple directions (multidimensional dMRI). If we consider each voxel as composed of multiple Gaussian compartments, then the signal from any q-space trajectory is fully specified by 
a rank-2 symmetric b-tensor. Thus, we can optimise over all q-space trajectories by optimising over all b-tensors. These are
defined by \cite{WESTIN2016}
\begin{equation}\label{eq:Btensor}
\boldsymbol{B} = \gamma\! \int_0^\tau\!\!\mathbf{q}(t')\! \otimes\! \mathbf{q}(t')\, dt', \quad \text{with} \  b = \Tr(\boldsymbol{B}) = B_{ii}, \quad \mathbf{q}(t)=\gamma \!\int_0^t \!\!\mathbf{g}(t')\,dt',
\end{equation}
where $\tau$ is the echo time, $b$ the conventional b-value or diffusion weighting, and $\mathbf{q}(t)$ is the time-dependent gradient waveform with gyro-magnetic ratio $\gamma$ and gradient $\mathbf{g}(t)$. We use the Einstein summation convention
. Figure \ref{fig:Tensor_Shapes} shows different b-tensors in a triangular diagram with the standard ones (linear, planar, and spherical) at the vertices. The eigenvalues define the shape and diffusion weighting (\textit{i.e.} size) and eigenvectors the orientation.

Multidimensional dMRI was proposed in \cite{WESTIN2016}, assuming an underlying diffusion tensor distribution, to disentangle orientation dispersion and microstructural anisotropy. 
We focus on applying multidimensional dMRI to biophysical modelling. The SM considers a \textit{stick} (or \textit{intra-neurite}) compartment with restricted diffusion, representing axons and, possibly, glial processes, where we assume the diffusion occurs only along one direction (\textit{i.e.} fibre orientation). 
These are embedded in a free but anisotropic space, represented by an \textit{extra-neurite} compartment where hindered diffusion is modelled  
as Gaussian with cylindrical symmetry.
Negligible water exchange between the compartments is assumed for typical experimental time scales. A \textit{fibre segment} is defined as a local bundle of parallel sticks with the extra-neurite space surrounding them. Voxels comprise a large number of fibre segments with equal microstructural properties, situated according to a fibre orientation distribution function (fODF)
 $\mathcal{P}(\mathbf{n})$. Thus, the dMRI signal 
 is the convolution over the unit sphere of the fibre response signal and the fODF:
\begin{equation}\label{eq:IntegratedKernel}
	S(\boldsymbol{B})=S_0 \int_{\mathbb{S}^2} {\cal K}(\boldsymbol{B};\hbu) \, {\cal P}(\hbu) \dif \hbu,
\end{equation}
with the response signal (also called kernel) being
\begin{equation}\label{eq:Kernel_MDE}
	{\cal K}(\boldsymbol{B};\hbu)=f_{\text{a}}\,e^{-D_{\text{a}}u_iu_j B_{ij}} + 
	f_{\text{e}}\, e^{-(D_{\text{e}}^\perp\delta_{ij}+\Delta_{\text{e}}u_iu_j)B_{ij}} +
	(1-f_{\text{a}}-f_{\text{e}})\,e^{-B_{ii}\,D_{\text{CSF}}},
\end{equation}
where $\hbu$ is the fibre segment's main axis, $f_a$ and $f_e$ the (mainly) $T_2$-weighted intra-neurite and extra-neurite volume fractions, $D_\text{a}$ the intra-neurite axial diffusivity, $D_\text{e}^\parallel$ and $D_\text{e}^\perp$ the extra-neurite diffusivities parallel and perpendicular to fibre's main axis, and $D_\text{CSF}$ the cerebrospinal fluid (free-water) diffusivity. 


Up to intermediate diffusion weightings (\textit{i.e.} $b<2.5ms/\mu m^2$), the dMRI signal can be accurately represented with a fourth order cumulant expansion 
\cite{KISELEV2007}. For a multidimensional dMRI acquisition this is given by
\begin{equation}\label{Eq:MultidimensionalCumulantFourthOrderD}
\log (S) = \log(S_0) - B_{ij} D_{ij} +\tfrac12 B_{ij}B_{k\ell} C_{ijk\ell},
\end{equation}
where $\boldsymbol{D}$ is the diffusion tensor and $\boldsymbol{C}$ is the second cumulant tensor, 
which with 
multiple Gaussian compartments coincides with the diffusion tensor covariance. 
The 
symmetric part of $\boldsymbol{C}$ is proportional to the well-known kurtosis tensor: $\bar{D}^2 W_{ijkl} = 3 C_{(ijkl)}$.


\section{Methods}

\subsection{Information Metric}

We are interested in the B-set that provides the most accurate and precise parameter estimates. To obtain this, we could compute numerically the estimation error for a certain B-set and get the one that minimises such error. Since this depends on the kernel and ODF parameters, it would have to be repeated over the full SM parameter space to integrate it numerically and remove this dependence. This makes its computation time-consuming and impractical for multidimensional optimisation. 
Many works use instead a surrogate to rank experimental designs, such as theoretical bounds for the variance of the estimators of the parameters of interest. 

We propose to maximise the \textit{information} about the parameters of interest provided by the B-set. This information is quantified here by the determinant of its FIM. Maximising the determinant of the FIM is equivalent to minimising the determinant of its inverse, 
which for unbiased estimators is analogous to minimising the generalised error variance \cite{SCHARF1991}. 
However, in a multimodal likelihood, the FIM does not depend on the number of modes or their distances, only on the sharpness of each mode. 
Multimodality confounds the estimation when only one set of measurements is available, increasing the error. 
For LTE acquisitions, the SM's likelihood is multimodal and results in degenerated estimators \cite{NOVIKOV2018}. This multimodality would not be penalised by any criterion based on the FIM, potentially wrongly selecting an optimal experimental setting including only LTE. 

To overcome this issue, we slightly change our target to optimise the acquisition for the estimation of the 4th order cumulant expansion tensors $\boldsymbol{D}$ and $\boldsymbol{C}$ (Eq. \ref{Eq:MultidimensionalCumulantFourthOrderD}) instead of the full signal. This presents interesting properties. First, this is a convex problem, thus unimodal. Second, the SM parameters can be fully 
determined from the cumulant tensors \cite{COELHO2019a}. Although the accuracy of the SM parameters is not necessarily monotonic regarding the accuracy of the cumulant tensors, the cumulant-optimal acquisition can be a good surrogate for the SM-optimal acquisition. In addition, the cumulant-optimal could be considered a generalised problem, providing also surrogates for more general models beyond the SM.

Considering that the signal-to-noise ratio (SNR) is $>2$, we can approximate MR Rician noise as Gaussian \cite{GUDBJARTSSON1995}. Then, the FIM for a B-set of K tensors $\{\boldsymbol{B}_k\}_{k=1,\ldots,K}$ is given by
\begin{equation}\label{eq:FIM_GaussianNoise}
	J(\{\boldsymbol{B}_k\};\mathbf{\Theta})_{ij} = 
    \sigma^{-2} \, 
    \sum_{k=1}^K 
    \frac{\partial S(\boldsymbol{B}_k;\mathbf{\Theta})}{
    	\partial \Theta_i} \, 
    \frac{\partial S(\boldsymbol{B}_k;\mathbf{\Theta})}{
    	\partial \Theta_j},
\end{equation}
where $\sigma$ is the noise standard deviation, and $\mathbf{\Theta}=[\boldsymbol{D},\boldsymbol{C}]$, thus, $i,j=1,\ldots,27$.


Since $\mathbf{J}(\{\mathbf{B}_k\};\mathbf{\Theta})$ depends not only on $\{\mathbf{B}_k\}$ but also on the tissue $\mathbf{\Theta}$, a tissue-independent metric was defined. We average the FIM by integrating it over the full parameter space $\cal{H}$ and dividing it by the integration volume:
\begin{equation}\label{eq:FIM_Integrated}
	\boldsymbol{\hat{J}}(\{\boldsymbol{B}_k\}) = \frac{1}{\text{vol}(\cal{H})}\int_{\cal{H}}  
    \mathbf{J}(\{\boldsymbol{B}_k\};\mathbf{\Theta}) \, d\mathbf{\Theta}.
\end{equation}
An advantage of the cumulant expansion 
is that 
Eq \ref{eq:FIM_Integrated} has a closed analytical solution and the result is rotationally invariant. Integration bounds for each tensor element were computed numerically to include the feasible SM parameters in the brain. We generated a grid of 300000 points in the physically plausible SM parameter space ($0\leq f\leq1$, positive diffusivities smaller than water diffusion, and a single Watson fODF) and computed the corresponding $\boldsymbol{D}$ and $\boldsymbol{C}$ tensors. We selected the ranges containing $90\%$ of the points. 
Our experiments show that the ranking of 
B-sets is not significantly affected by modifying these ranges.

\subsection{Optimisation strategy}

Considering an acquisition with $K$ measurements, $\{\mathbf{B}_k\}$ has $6K$ degrees of freedom if we do not impose constraints on the B-set. Due to the high dimensionality of the problem and multiple local minima, a hybrid two-step optimisation strategy was used. The first step consisted of stochastic optimisation, where the Self-Organising Migrating Algorithm (SOMA) \cite{ZELINKA2004} was selected. 
The second step included a local search with a gradient-descend method that used the output of the stochastic optimisation as the initial condition. This combined the robustness against local minima of stochastic optimisation and the rapid convergence, once the neighbourhood of the global optimum was found, of greedy approaches. We tested the robustness of our hybrid strategy in 
Ackley's function \cite{ACKLEY1987} for 50, 100, and 200 dimensions. Our hybrid approach found consistently the global optimum (results not shown).

\subsection{Experiments}
Since b-tensors are positive semidefinite, we parametrised them with their eigenvalues $\lambda_1, \lambda_2, \lambda_3$ and eigenvectors $\hat{\mathbf{v}}_1, \hat{\mathbf{v}}_2, \hat{\mathbf{v}}_3$, and restricted the search space to $\lambda_1 + \lambda_2 + \lambda_3\geq 0$. We considered a constant SNR, independent of the diffusion weighting, which was limited between $0-2 ms/\mu m^2$. For our experiments we considered a B-set of 60 tensors. Due to the high dimensionality of the problem, we applied two 
constraints on the B-set, progressively increasing the searching volume in the measurement space. The first one (C1), fixes the trace ($\lambda_{1,i} + \lambda_{2,i} + \lambda_{3,i} = b_i$) and eigenvectors of each b-tensor and only estimates their shapes. Measurements were grouped into two shells of $1-2 ms/\mu m^2$ with 30 b-tensors each and eigenvectors were distributed uniformly in the hemisphere
. The second (C2), only fixed the eigenvectors, leaving individual shapes and sizes free to vary, $0\leq \lambda_{1,i} + \lambda_{2,i} + \lambda_{3,i} \leq b_\text{max}$. These constraints resulted in $2K=120$ (C1) and $3K=180$ (C2) free parameters. SOMA was run 4 times with a population of 60 times the number of free parameters and 1500 migrations.


\section{Results}

\begin{figure}[htbp]
\centerline{\includegraphics[scale=.28]{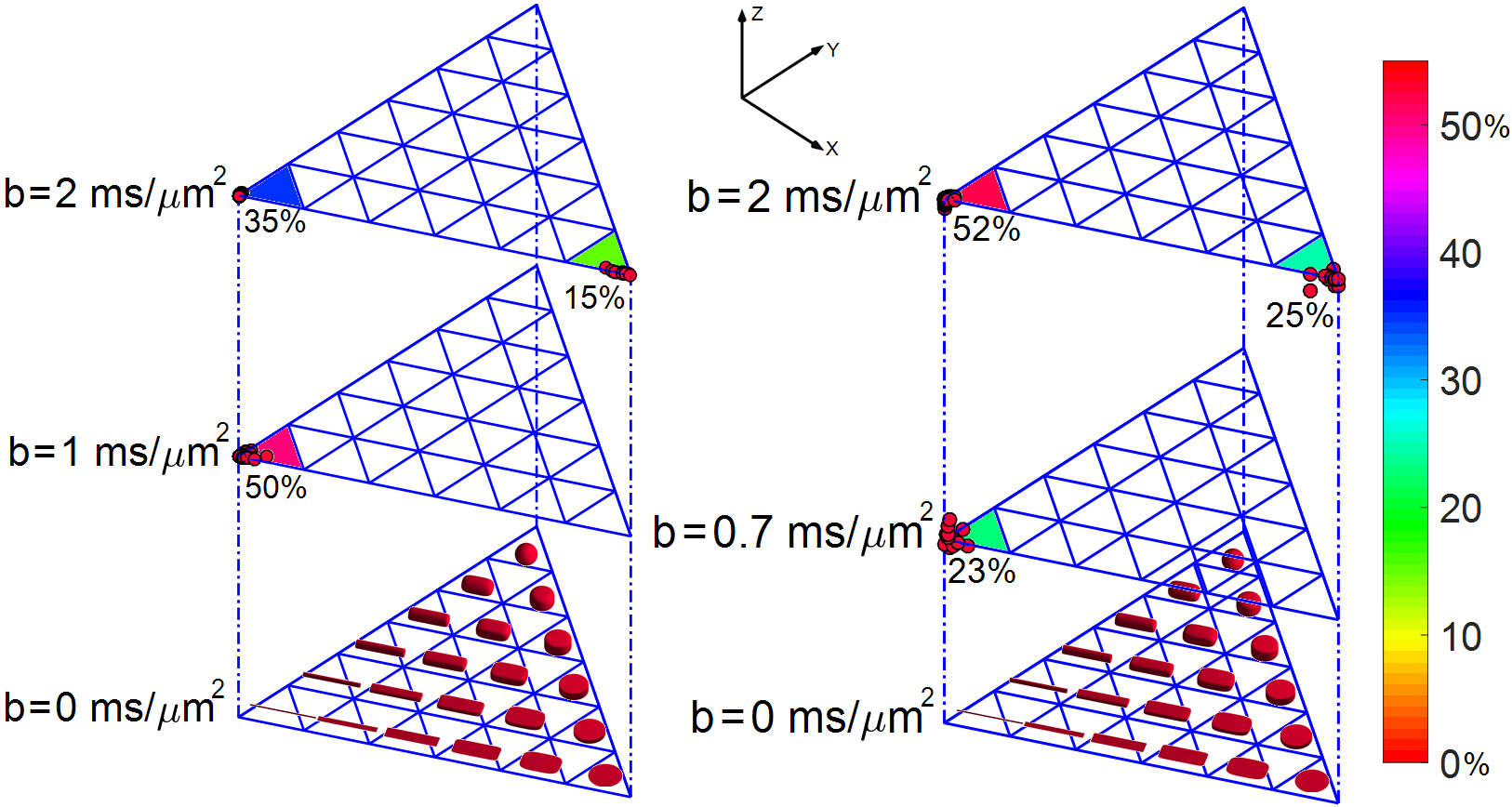}}
\caption{Optimal experimental designs considering each set of constraints (C1 left, C2 right). Individual dots represent the shapes of each tensor in each B-set (along XY plane in a triangular grid based on Fig. \ref{fig:Tensor_Shapes}), with their corresponding sizes (along Z axis). To appreciate proportions we added the percentages of points in each cluster regarding the total number of points and coloured the triangle containing them.}
\label{fig:Optimal_Protocols}
\end{figure}

Optimal configurations are very similar for both sets of constraints. Figure \ref{fig:Optimal_Protocols} shows resulting B-sets. The shapes of the b-tensors are distributed between linear and planar encoding only. There are no spherical tensors. This is unsurprising since STE measurements are only sensitive to the traces of $\boldsymbol{D}$ and $\boldsymbol{C}$, while LTE and PTE each \textit{excite} 21 of the 27 independent parameters. However, there are no other mixed shapes either. The optimal proportion of LTE data is around $75\%-85\%$, depending on $tr(B)$ being fixed or not. An interesting result is that in the C2 optimisation, where the traces are a free parameter, measurements group into two shells with unequal number of measurements and only LTE measurements on the lower shell.


To assess how good are these optima as surrogates for the SM, we performed a Monte Carlo experiment comparing the optimal protocol from C2 against a balanced combination of LTE and PTE data \cite{COELHO2019a}. We computed the synthetic signals from 10000 random voxels corresponding to both acquisitions, added Rician noise (SNR=100) and computed the estimation error. Signals were generated both with Eq. \ref{Eq:MultidimensionalCumulantFourthOrderD} and Eq. \ref{eq:IntegratedKernel} considering a double-Watson ODF and a null CSF fraction. The improvement in the accuracy of each parameter by using C2 over the naive combination was $14\%$ in the estimation of $\boldsymbol{D}$ and $\boldsymbol{C}$, and $7\%$ in the estimation of the double-Watson SM. 


\section{Discussion}

This work proposed a novel framework to compute the optimal B-set that reduces the error in the estimation of the second cumulant of the dMRI signal, and consequently, improves the parameter estimation of biophysical models. We defined a metric to quantify the information provided by each B-set, based on the computation of an averaged Fisher information matrix. Thus, the framework only depends on the acquisition 
since the metric is averaged over the entire tissue parameter space.  
Currently no work in the literature attempts to optimise a multidimensional dMRI acquisition for biophysical models.

By using the proposed proportions of LTE and PTE measurements, and their corresponding diffusion weightings, the estimation error of each parameter was reduced by $7\%$ on average, regarding naively combining LTE and PTE for a SM with a double-Watson ODF. Although this might be a suboptimal acquisition for the SM because the B-set was optimised for $\boldsymbol{D}$ and $\boldsymbol{C}$, it shows a cost-free improvement. The confirmation that only LTE and PTE data are required for an optimal acquisition is interesting. Future work will include comprehensive \textit{in silico} experiments including different number of measurements and experiments with different fODFs. The average improvement might also increase if we are mainly interested in optimising the estimation of a specific parameter of interest. Within this framework, we will also explore modifications of the information metric to ensure that the B-set is closer to the SM's global optimal.

One limitation of this work is that we computed the FIM with the fourth-order cumulant expansion instead of the SM analytical signal. However, up to $\text{b}_{max}=2.5 ms/\mu m^2$ this is 
a reasonable approximation \cite{KISELEV2007}, and therefore has the potential to be a good enough surrogate for experimental design in this range. Additionally, this is a more general result, which is possibly a good surrogate also for biophysical models beyond the SM. 
Since we considered optimistic SNRs, the noise model was approximated as Gaussian. Integrating the FIM over the cumulant parameter space was performed with some approximations, but these do not seem to interfere when ranking different protocols. Since the optimisation is high dimensional and with multiple local minima, further improvements on the optimisation strategy may improve the results. However, 
our hybrid approach showed robust estimations of global optima in a toy function of similar dimensionality and complexity. 
Finally, relaxing the constraints will likely improve the results, such as releasing the eigenvectors or considering different SNRs for each diffusion weighting. We will consider them in future work.



\section{Conclusions}
We proposed a framework to compute the set of b-tensors that maximises the microstructural information in the multidimensional dMRI acquisition. This can help to reduce the acquisition time aimed at estimating the \textit{Standard Model} or any given biophysical model, or reduce the error in the estimated parameters. 
The framework is based on maximising the determinant of the Fisher information matrix averaged over the expected values in the parameter space.
Our results were consistent between different constraints considered, showing in both that the optimal sampling was reached by combinations of LTE and PTE data with two diffusion weightings.

\section*{Acknowledgements}

This work has been supported by the OCEAN project (EP/M006328/1) and MedIAN Network (EP/N026993/1) both funded by the Engineering and Physical Sciences Research Council (EPSRC) and the European Commission FP7 Project VPH-DARE@IT (FP7-ICT-2011-9-601055).

\bibliographystyle{ieeetr}
\bibliography{Coelho_bibliography_2019_01_25.bib} 

\begin{thebibliography}{10}

\bibitem{KISELEV2016}
V.~G. Kiselev, ``Fundamentals of diffusion {MRI} physics,'' {\em {NMR} in
  Biomedicine}, vol.~30, pp.~1--18, 2017.

\bibitem{ASSAF2008b}
Y.~Assaf, ``Can we use diffusion {MRI} as a bio-marker of neurodegenerative
  processes?,'' {\em BioEssays}, vol.~30, no.~11-12, pp.~1235--1245, 2008.

\bibitem{ASSAF2004a}
Y.~Assaf, T.~Blumenfeld-Katzir, Y.~Yovel, and P.~J. Basser, ``New modeling and
  experimental framework to characterize hindered and restricted water
  diffusion in brain white matter,'' {\em Magnetic Resonance in Medicine},
  vol.~52, pp.~965--978, 2004.

\bibitem{NOVIKOV2018}
D.~S. Novikov, J.~Veraart, I.~O. Jelescu, and E.~Fieremans,
  ``Rotationally-invariant mapping of scalar and orientational metrics of
  neuronal microstructure with diffusion {MRI},'' {\em NeuroImage}, vol.~174,
  pp.~518 -- 538, 2018.

\bibitem{JELESCU2015b}
I.~O. Jelescu, J.~Veraart, E.~Fieremans, and D.~S. Novikov, ``Degeneracy in
  model parameter estimation for multi-compartmental diffusion in neuronal
  tissue,'' {\em NMR in Biomedicine}, vol.~29, pp.~33--47, 2016.

\bibitem{COELHO2019a}
S.~{Coelho}, J.~M. {Pozo}, S.~N. {Jespersen}, D.~K. {Jones}, and A.~F.
  {Frangi}, ``Resolving degeneracy in diffusion {MRI} biophysical model
  parameter estimation using double diffusion encoding,'' {\em Magnetic
  Resonance in Medicine}, vol.~82, pp.~395--410, 2019.

\bibitem{REISERT2019}
M.~Reisert, V.~G. Kiselev, and B.~Dhital, ``A unique analytical solution of the
  white matter standard model using linear and planar encodings,'' {\em
  Magnetic Resonance in Medicine}, vol.~81, pp.~3819--3825, 2019.

\bibitem{WESTIN2016}
C.-F. Westin, H.~Knutsson, O.~Pasternak, F.~Szczepankiewicz, E.~\"Ozarslan,
  D.~van Westen, C.~Mattisson, M.~Bogren, L.~J. O'Donnell, M.~Kubicki,
  D.~Topgaard, and M.~Nilsson, ``q-space trajectory imaging for
  multidimensional diffusion {MRI} of the human brain,'' {\em NeuroImage},
  vol.~135, pp.~345--362, 2016.

\bibitem{TOPGAARD2017}
D.~Topgaard, ``Multidimensional diffusion {MRI},'' {\em Journal of Magnetic
  Resonance}, vol.~275, pp.~98--113, 2017.

\bibitem{KISELEV2007}
V.~G. Kiselev and K.~A. Il'yasov, ``Is the “biexponential diffusion”
  biexponential?,'' {\em Magnetic Resonance in Medicine}, vol.~57, no.~3,
  pp.~464--469, 2007.

\bibitem{SCHARF1991}
L.~L. Scharf, {\em {Statistical} {Signal} {Processing}: Detection, Estimation,
  and Time Series Analysis}.
\newblock Addison-Wesley Publishing Company, 1991.

\bibitem{GUDBJARTSSON1995}
H.~Gudbjartsson and S.~Patz, ``The {Rician} distribution of noisy {MRI} data,''
  {\em Magnetic Resonance in Medicine}, vol.~34, no.~6, pp.~910--914, 1995.

\bibitem{ZELINKA2004}
I.~Zelinka, ``Soma --- self-organizing migrating algorithm,'' in {\em New
  Optimization Techniques in Engineering}, ch.~7, pp.~167--217, Springer Berlin
  Heidelberg, 2004.

\bibitem{ACKLEY1987}
D.~Ackley, {\em A connectionist machine for genetic hillclimbing}.
\newblock Kluwer Academic Publishers, 1987.

\end{thebibliography}

\end{document}